\documentclass[5p,twocolumn]{elsarticle}
\usepackage{lineno,hyperref}
\modulolinenumbers[5]
\usepackage{graphicx}
\usepackage{dcolumn}
\usepackage{bm}
\usepackage{amsmath}

\begin{document}
\begin{frontmatter}
  
\title{Quadrupole deformation signatures in elastic electron scattering from 
oriented odd-A nuclei}

\author{P.~Sarriguren}
\address{Instituto de Estructura de la Materia (IEM), CSIC, Serrano 123,
  E-28006 Madrid, Spain}
\ead{p.sarriguren@csic.es}

\date{\today}

\begin{abstract}
Elastic electron scattering from oriented odd-$A$ axially deformed nuclei is 
studied in the plane-wave Born approximation. The nuclear structure is described
within a microscopic selfconsistent Skyrme deformed Hartree-Fock approximation 
with pairing correlations. The interference form factors between monopole and 
quadrupole Coulomb terms that characterize the nuclear response with aligned 
nuclear targets are shown to increase or decrease the unpolarized cross-section,
depending on the sign of the quadrupole deformation. This feature provides 
valuable information on the nuclear deformation that can be used as a signature 
of the oblate or prolate character of the nuclear shape. Some selected cases 
of nuclei with different spins are presented that exemplify the scope of the 
method.
\end{abstract}

\begin{keyword} elastic electron scattering, axially deformed nuclei, selfconsistent mean field, 
charge form factors
\end{keyword}

\end{frontmatter}

\section{Introduction}

Nuclear deformation is a crucial piece of information to fully characterize the
bulk properties of nuclei \cite{bm,ring}. The shape of the nucleus not only 
determines a large variety of its properties, such as rotational spectra, 
electromagnetic moments and transition strengths, but it is also essential to
understand nuclear decays and reactions with important consequences in other 
fields. Indeed, the implications of the nuclear shapes extend beyond the realm 
of nuclear physics itself with impact in nuclear astrophysics and particle 
physics, where nuclear deformation has been shown to be an important issue to 
properly understand nuclear reactions and decays in stellar environments 
\cite{sarri-stellar}, as well as rare processes such as the double beta 
decay \cite{2bb-def1,2bb-def2,2bb-def3}. 

A complete description of the nuclear shape requires in principle a large number
of parameters to account for all degrees of freedom. However, a first step to 
characterize the shape of the nucleus beyond sphericity can be made by assuming 
axial symmetry and quadrupole deformation. This approximation is still valid 
to describe properly enough a large variety of the existing nuclei. Neglecting 
other degrees of freedom, such as triaxiality or higher multipolarities, will 
not be important for the purpose of this work. The shape of a nucleus with such 
symmetry is commonly characterized by the quadrupole deformation parameter 
($\beta_2$) whose sign indicates whether the axis of symmetry is larger 
(prolate) or smaller (oblate) than the perpendicular axes. In the former 
case  $\beta_2$ is positive, in the latter negative.

The nuclear deformation can be theoretically calculated as the shape 
configuration that minimizes the energy. However, experimental information 
on the quadrupole deformation parameter and especially its sign, is still 
limited \cite{stone}. In particular, electric quadrupole transition 
probabilities $B(E2)$ from the first excited $2^+$ states in even-even nuclei 
have been systematically used to get information about the magnitude of 
$\beta_2$ \cite{pritychenko}, but not about its sign because of the quadratic 
dependence on $\beta_2$.  

More detailed information about the nuclear structure and specifically about the 
nuclear shape is provided by Coulomb excitation experiments that simultaneously 
measure excitation energies and cross-sections \cite{coulex1,coulex2}. Analysis 
of the $\gamma$-rays in the measured spectrum provides the transition energies
between bound states, while the yield of $\gamma$-rays with a given energy 
measures the Coulomb excitation cross-section, which can be related with the 
electromagnetic transition matrix elements. The development of analytical 
techniques now available (GOSIA codes) makes it possible to extract detailed 
information about static moments and relative signs and magnitudes of 
non-diagonal matrix elements, which allow to determine both the centroids and 
fluctuation widths of the quadrupole shape degrees of freedom.

Similarly, because the sensitivity of the $\beta$-strength distribution to 
deformation, $\beta$-decay experiments have also been proposed as a 
complementary tool to obtain information about the nuclear shape of unstable 
nuclei \cite{nacher,sarri-beta}. 

An alternative method to get information on the sign of the quadrupole 
deformation that has not been sufficiently explored yet is to use the 
sensitivity of the electron scattering experiments to the nuclear charge 
density distribution and, as a result, to the nuclear deformation. The 
electron scattering cross-section contains the information of the full 
charge density distribution of the nucleus, which  appears through the 
charge form factors \cite{e-scatt1,e-scatt2,e-scatt3,e-scatt4}.

The relationship between the charge form factors and the properties that
characterize the charge density distributions has been studied within
different approaches that include the Shell Model \cite{brown,radhi}, 
as well as relativistic 
\cite{wang-prc71,roca-prc78,roca-prc87,liu-prc95,liu-prc96,liang-prc98,liu-jpg46}
and nonrelativistic \cite{richter,anton_2005,sarri_prc76,yao-prc91,wang-jpg47} 
selfconsistent mean-field models. All these models show clear correlations
between them. In particular, the value of the momentum transfer at the first
minimum of the charge form factor is related with the r.m.s. radius of the 
charge distribution \cite{wang-prc71,roca-prc78,roca-prc87,sarri_prc76},
whereas the height of the second maximum is related with the surface 
diffuseness of the charge distribution \cite{sarri_prc76,yao-prc91,friedrich}.

Isotopic and isotonic chains have been studied in the above mentioned works
to stress different aspects of nuclear structure. Isotopic chains are useful
to study the effect of neutrons on the charge density that becomes more 
diluted as the number of neutrons increases. Similarly, the effect of the outer 
protons can be analyzed in isotonic chains and details of the proton wave 
functions can be studied. An upward and inward shift of the diffraction 
minima of the charge form factors is found with increasing nucleon number 
in both type of chains.

The Coulomb form factors for deformed nuclei can be expressed in terms of 
multipoles $C \lambda$ \cite{e-scatt3,moya}, which are sensitive to the 
different components of a multipolar decomposition of the deformed charge 
density distribution. In particular, the $C0$ component would contain the 
information on the dominant spherical component, whereas the $C2$ component 
would contain the information on the quadrupole nuclear deformation. 
However, in traditional electron scattering, where unpolarized electron 
beams are scattered from unpolarized nuclear targets, the total charge form 
factor appears as an incoherent sum of multipoles squared. Therefore, the 
information sensitive to the nuclear deformation is hidden because of two 
main reasons. First, the contributions from higher multipolarities to the 
cross-section rapidly become smaller with increasing multipolarity. Second, 
only the square of $C2$ multipoles contributes and this prevents extracting 
information about the sign that determines the oblate or prolate character 
of the deformation. 

The combined effects of both quadrupole deformation and r.m.s. radii on the 
diffraction minima of the charge form factors have also been studied 
\cite{liu-prc95,liu-prc96,liang-prc98,liu-jpg46,wang-jpg47}.
However, because of the above reasons, information on the nuclear deformation 
is difficult to be disentangled. For example, it has been shown \cite{liu-prc96} 
that in ordinary electron scattering $C2$ contributions manifest significantly 
only at the $C0$ diffraction minima, but unfortunately, this occurs at 
relatively large momentum transfer, where the form factors are small and 
difficult to measure. 

One possible way to solve this problem is the use polarization degrees of 
freedom in the electron scattering from nuclei \cite{moya,weigert,donnelly}. 
In this case, new observables appear that contain interference terms between 
the various multipoles that can be isolated experimentally with proper choices 
of the directions of both scattered electrons and nuclear polarizations. 
In particular, interference terms between longitudinal multipoles $C0/C2$ 
appear in the case of scattering from unpolarized electrons from oriented 
nuclei and they will be the focus in this work. These interference terms are 
much larger than the $C2$ squared terms that characterize the unpolarized 
cross-sections and involve phases sensitive to the quadrupole sign that 
increase or decrease significantly the unpolarized cross-sections.
The essential point is that new combinations of the form factors (other than 
incoherent sums of squares) can be measured by varying the polarization 
direction of the nucleus.

In the simplest realization of these experiments, polarized internal targets 
are located to intercept the circulating electrons in storage rings. The 
high currents achieved in these storage rings allow the use of thin targets 
that can be more easily polarized, reaching luminosities comparable with 
those in conventional experiments \cite{exp_pol_1,exp_pol_2}.
New possibilities with unprecedented polarization quality of both electron 
and ion beams are also expected in the new generations of facilities, 
in particular, in the Electron-Ion Collider at Brookhaven National 
Laboratory \cite{eic}.

Most of the scattering experiments in the past were carried out with stable 
nuclei. In this work we focus on examples among them. However, this could
change soon with the setup of collider experiments with electrons and 
radioactive ion beams. The SCRIT project at RIKEN \cite{unstable,scrit_xe} 
and the ELISe project planned for the GSI-FAIR \cite{unstable,anton_nima}
are examples of these new facilities.

In the next section the theoretical formalism used to describe the reaction
mechanism and the nuclear structure of the electron scattering from deformed
aligned odd-$A$ nuclei is presented. Section 3 contains the results of this
study on the examples of $^{23}$Na, $^{25}$Mg, and $^{59}$Co. The main 
conclusions are shown in the last section.

\section{FORMALISM}

The reaction mechanism used in this work is based on the plane wave Born
approximation (PWBA) that assumes one-photon exchanged and plane waves 
for the electrons. In addition to the numerical advantage, PWBA relates 
directly the charge form factors with the Fourier transforms of the charge 
matrix elements and the physical interpretation is straightforward. 
Coulomb distortion effects evaluated from the distorted wave Born 
approximation (DWBA) will be needed for a final comparison with experiment. 
They can be included within a phase-shift analysis \cite{yennie}. 
However, the main effects of Coulomb distortion are the filling of the
diffraction minima and a shift of their locations. These effects are small at
low momentum transfer ($q\leq 1.2$ fm$^{-1}$), where the effects from target
orientation are already significant.

The theoretical formalism for inclusive electron-nucleus scattering involving
polarization degrees of freedom in both beams and targets has been developed
elsewhere \cite{weigert,donnelly} and has been particularized to the case of
deformed nuclei in Refs. \cite{moya,garrido}. Here, we only summarize the most
relevant results regarding the nuclear longitudinal $(L)$ response functions
(also known as charge or Coulomb responses) for experiments dealing with 
polarized targets without measuring electron polarizations. Similar results 
can be obtained in the case where the target is not polarized, but the 
polarization of the final nucleus is measured. The longitudinal response 
can always be separated from the transverse one that involves electric and 
magnetic contributions by a Rosenbluth analysis, using the dependence of 
the different weighting factors on the electron kinematics.

Following the notation in Ref. \cite{moya}, the electron scattering 
cross-section in PWBA from deformed oriented nuclei with a ground state 
$I_i^{\pi_i}$ characterized by spin $I_i$ and parity $\pi_i$ to a final 
state $I_f^{\pi_f}$ is given by

\begin{equation}
  \left. \frac{d\sigma}{d\Omega} \right| _{I_i^{\pi_i}\rightarrow I_f^{\pi_f}} = 
\sigma_{\rm tot} (\theta', \phi') = 
Z^2 \sigma_M f_{\rm rec}^{-1}
\left[ \sigma_0 + \sigma_{\rm al} (\theta', \phi') \right] \, ,
\end{equation}
where $\sigma_M$ is the Mott cross-section, $f_{\rm rec}$ the recoil factor, 
and the angles $(\theta', \phi')$ represent the polarization direction of 
the target relative to the system determined by the momentum transfer
$\vec{q} =\vec{k_i}-\vec{k_f}$. Figures showing the relationship between the
target polarization direction, the  $\vec{q}$-system and the laboratory
system can be found in Ref. \cite{moya}.

The cross-section $\sigma_0$ does not depend on polarizations and is given by

\begin{equation}
\sigma_0 = V_L \left| F_L(q) \right| ^2 
+ V_T \left| F_T(q) \right| ^2 \, .
\end{equation}
$V_L$ and $V_T$ are kinematical factors given by

\begin{equation}  
V_L=(Q^2/q^2)^2\ , \quad  V_T=\tan^2 (\theta_e/2) - (Q^2/q^2)/2\, ,
\end{equation}
where $Q^{\mu}=(\omega,\vec{q})$ is the four-momentum transferred to the nucleus
in the process in which an incoming electron with  four-momentum
$k_i ^{\mu}  =(\varepsilon_i,\vec{k_i})$ is scattered through an angle $\theta_e$
to an outgoing electron with  four-momentum
$k_f ^{\mu}  =(\varepsilon_f,\vec{k_f})$, with
$\omega=\varepsilon_i -\varepsilon_f$ and $\vec{q}=\vec{k_i}-\vec{k_f}$.
$F_L$ and $F_T$ are the longitudinal and transverse form factors that 
contain the nuclear structure information in terms of charge ($L$) and electric
and magnetic ($T$) multipoles. These contributions can be isolated with proper 
choices of the kinematical variables using Rosenbluth separation methods. 
The focus here is 
on the Coulomb response $F_L(q)$ that can be written in terms of the charge 
multipole form factors $F^{C\lambda}(q)$

\begin{equation}
  \left| F_L(I_i,I_f;q) \right| ^2 \, = \sum _ {\lambda \ge 0}
  \left| F^{C\lambda}(I_i,I_f;q)
  \right| ^2 \, .
\end{equation}
The various multipole contributions add incoherently and share the same factors
in the cross-section, making it impossible to separate them kinematically.
The multipole charge form factors are given by
\begin{equation}
  F^{C\lambda}(I_i,I_f;q) = \frac{\sqrt{4\pi }}{Z}\langle I_f || \hat T ^{C\lambda}(q)
  || I_i \rangle /\sqrt{2I_i+1}\, ,
\end{equation}
where the Coulomb multipole operators are
\begin{equation}
\hat T ^{C\lambda} _\mu (q) = i ^{\lambda} \int d{\bf R} j_{\lambda}(qR) 
Y^{\mu} _\lambda (\Omega_R) \hat \rho ({\bf R}) \, ,
\end{equation}
in terms of the nuclear charge operator $\hat \rho ({\bf R})$.

Similarly, $\sigma_{\rm al}$ is independent of projectile polarizations and
depends on the orientation of the target.

\begin{align}
\sigma_{\rm al} (\theta', \phi')   =  \sum_{\ell =even>0} & \alpha_{\ell}^{I_i} 
 [ P_{\ell} (\cos \theta') \left( -V_L   F_L^{\ell} + V_T F_T^{\ell} \right) 
\nonumber \\
&
+ P_{\ell}^2 (\cos \theta') \cos (2\phi') V_{TT}  F_{TT}^{\ell} \nonumber \\
& 
+ P_{\ell}^1 (\cos \theta') \cos \phi' V_{TL}  F_{TL}^{\ell} ]
 \, .
\end{align}
$V_{TT}$ and $V_{TL}$ are kinematical factors whose explicit expressions can be
found in Ref. \cite{moya}. The terms $TT$ and $TL$ are not considered
further in this work because they do not contribute to the cross-section
when the nuclei are oriented in the $\theta'=0$ direction, which will be the
only case considered here.

The structure functions $F$'s contain different interference terms that can be 
separated using the dependence on the polarization direction $(\theta',\phi')$, 
the statistical factors  $\alpha_{\ell}^{I_i}$, and the dependence on the 
scattering angle $\theta_e$. In particular, the interest of this work focuses on the 
interference between longitudinal multipoles that are contained in the term  
$F_L^{\ell}$.

The statistical tensors describe the orientation of the target, reflecting 
that  the $2I_i+1$ substates are not equally populated,

\begin{equation}
\alpha_{\ell}^{I_i} = \sum_{M_i} P(M_i) \langle I_i M_i \ell 0 | I_i M_i \rangle \, ,
\label{stat}
\end{equation}
where $P(M_i)$ are the occupation probabilities of the target substates 
$|I_iM_i>$ along the polarization direction $(\theta', \phi')$. They are 
defined so that for $\ell =0$ one has $\alpha_{\ell =0}^{I}=1$ regardless of 
the state of polarization. For unpolarized targets where $P(M_i)=1/(2I_i+1)$, 
one has $\alpha_{\ell}^{I}=\delta_{\ell,0}$. In the case of aligned nuclei, 
$P(M_i)= P(-M_i)$, all the statistical tensors with odd values of $\ell$ 
vanish. In those cases the use of polarized electrons is irrelevant.

The form factors $ F_L^{\ell} $ contain interference terms 
between the different Coulomb multipoles:

\begin{equation}
F_L^{\ell} (I_i,I_f;q) = \sum _{\lambda, \lambda'} X(\lambda, \lambda', I_i, I_f, \ell )
F ^ {C\lambda}(I_i,I_f;q) F ^ {C\lambda'}(I_i,I_f;q)\, ,
\label{interfff}
\end{equation}
where  $X(\lambda, \lambda', I_i, I_f, \ell )$ are geometrical factors 
involving 3j- and 6j- Wigner coefficients \cite{moya},

\begin{equation}
X = (-1)^{I_i+I_f+1} {\bar I_i}{\bar \lambda}
{\bar \lambda '}{\bar \ell}^2 
\left(
        \begin{array}{ccc}
        \lambda   &  \lambda '   &  \ell  \\
        0 &  0 &  0
        \end{array}
\right) 
\left\{
        \begin{array}{ccc}
        I_i   &  I_i   &  \ell  \\
        \lambda &  \lambda ' &  I_f
        \end{array}
\right\} \, ,
\label{eq-x}
\end{equation}
with $\bar a=\sqrt{2a+1}$.

The transition multipole charge form factors $F^{C\lambda}(I_i,I_f;q)$ corresponding to 
a given transition $I_ik \rightarrow I_fk$ within a rotational band characterized
by the spin projection along the symmetry axis $k$, can be written up to lowest
order in angular momentum \cite{moya} in terms of intrinsic form factors

\begin{equation}
F^{C\lambda}(I_i,I_f;q) = \langle I_i k \lambda 0 | I_f k \rangle {\cal F} ^{C\lambda}(q) \, .
\label{intrinsic}
\end{equation}
They are given by

\begin{equation}
{\cal F} ^{C\lambda} (q) = i^{\lambda} \sqrt{\frac{4\pi}{2\lambda+1}}  
\int_0^\infty R^2dR \rho_\lambda(R) j_\lambda (qR)\, ,
\end{equation}
where  $\rho_\lambda(R)$ are the multipole components of an expansion of
the density  $\rho ({\bf R})$ in Legendre polynomials,

\begin{equation}
  \rho_{\lambda}(R)= (2\lambda +1) \int _{0}^{+1}  P_{\lambda}(\cos \theta)
  \rho (R \cos \theta , R \sin \theta) d(\cos \theta) \, ,
\end{equation}
Note that in axially symmetric nuclei the density depends only on two coordinates,
$R\cos \theta$ and $ R\sin \theta $.

In all the calculations in this work, center of mass corrections are considered 
in the harmonic-oscillator approximation, including a factor $\exp [q^2/(4A^{2/3})]$.
Nucleon finite size effects are included as a sum of monopoles parametrized in 
Ref. \cite{simon} for the proton and by the difference of two Gaussians 
\cite{chandra} for the neutron.

The nuclear structure used to describe the nuclear densities needed for the 
calculations of the different multipole charge form factors are obtained 
within a deformed Skyrme (SLy4) Hartree-Fock (HF) with pairing correlations 
in the BCS approximation, as described in Ref.  \cite{vautherin}.
In this formalism, the nuclear density is given by

\begin{equation}
\rho ({\bf R})= 2\sum_i v_i^2 \left| \Phi_i({\bf R})\right| ^2 \, ,
\end{equation}
in terms of the occupation probabilities $v_i^2$ and the single-particle 
Hartree-Fock wave functions $\Phi_i$, which are expanded into the eigenstates 
of an axially deformed harmonic oscillator potential \cite{vautherin}.
The quadrupole deformation $\beta_2$ is obtained selfconsistently from the
intrinsic quadrupole moment $Q_0$ and the mean square radius $<R ^2>$, both
calculated microscopically in terms of the density,

\begin{eqnarray}
  \beta_2 &=& \sqrt{\frac{\pi}{5}}\frac{Q_0}{A<R ^2>}\, , \nonumber \\
  Q_0 &=& \sqrt{16\pi/5}\int \rho(\vec{R})R^2 Y_{20}(\Omega_R) d\vec{R}\, , \nonumber \\
  \langle R^2 \rangle &=&  \frac{ \int R^2 \rho(\vec{R}) d\vec{R}}
  {\int \rho(\vec{R}) d\vec{R}} \, .
\label{beta2}
\end{eqnarray}

This formalism to describe axially deformed nuclei has been used in the past 
for the calculation of both longitudinal and transverse form factors
\cite{kowalski,berdi,graca,sarri_ff,nk,ff1,ff2}.

The main objective of the present work is the study of the interference 
between the monopole $C0$ and quadrupole $C2$ charge form factors, as 
contained in Eq. (\ref{interfff}). 
For even-even rotators $(I=k=0)$ there is no possibility of target orientation.
In this case, information on the $C2$ multipoles can only be extracted as 
$|F^{C2}|^2$ from ordinary inelastic transitions to the $2^+$ excited states 
that prevents getting  information about the sign of deformation.
Things are different for odd-$A$ rotors that can be aligned.
In this case, only elastic scattering will receive $C0$ contributions.
Therefore, the focus here is on the elastic scattering from odd-$A$ aligned
nuclei with $I=k \ge 3/2$, where $C0$ and $C2$ multipoles are 
involved and their interference will contribute to the cross-section.
In the case of elastic scattering $I_i=I_f$, the notation of
the charge multipoles is simplified to
$F^{C\lambda} (I_i=I_f,I_f;q) \equiv F^{C\lambda}_{I_f}(q)$.

Nuclei with $I=k=1/2$ cannot be aligned because they could not be distinguished
from the unpolarized case. In addition, elastic scattering from them will not 
receive $C2$ contributions, whereas inelastic transitions will not be sensitive 
to the $C0$. Therefore, for $k_i=1/2$ nuclei, only the special case where the 
ground state is given by $I_i=3/2$, which is possible because of Coriolis 
effects, will receive $C0/C2$ interferences. This case has already been 
discussed in Refs. \cite{moya,garrido}.

The key point is that by measuring both the unpolarized and the polarized 
cross-sections, one can see whether the target polarization increases or 
decreases the former, being a model independent signature of the oblate
or prolate character of the deformed target.

The form factors containing interference terms can be experimentally 
separated with measurements of the cross-sections at $\theta' =0$,
that is, with nuclei oriented in the  $\vec{q} =\vec{k_i}-\vec{k_f}$
direction

\begin{equation}
  \sigma_{\rm tot} (\theta'=0,\phi') -\sigma_0 = \sum_{\ell} \alpha_{\ell}^{I_i}
  \left( -V_L   F_L^{\ell} + V_T F_T^{\ell} \right)
\end{equation}
and using the kinematical dependence of the $V_L$ and $V_T$ factors, the 
contribution from  $F_L^{\ell}$ containing the interference of different 
charge multipoles can be isolated and studied separately.
The focus here is on the $C0/C2$ interferences that appear in the
$\ell=2$ terms,

\begin{equation}
\sigma _{\rm al} = - \alpha_2  ^{I_i} P_2 (\cos \theta') V_L F_L  ^{\ell=2}\, .
\label{k32}
\end{equation}
The $X$ coefficient in Eqs. (\ref{interfff}) and (\ref{eq-x}), in the elastic 
case $I_i=I_f=k$, and for $\ell=2$ involving the $C0/C2$ interference terms is 
given by $X=-2\sqrt{5}$, independent of the spin values. Then,

\begin{equation}
F_L^{\ell=2}(I_f;q) = -2\sqrt{5} F_{I_f}^{C0}(q) F_{I_f}^{C2}(q) \, .
\end{equation}
This form factor contains information about the relative phase of $C0$ and $C2$
multipole form factors and therefore, about the sign of the quadrupole charge
multipole $\rho_2$. This goes beyond the possibilities of ordinary electron
scattering, where the Coulomb contribution is given by

\begin{equation}
\sigma _{\rm 0} = V_L |F_L|^2 =  V_L \left( |F_{I_f}^{C0}|^2 + |F_{I_f}^{C2}|^2 + 
{\cal O} (C\lambda \ge 4) \right)\, .
\end{equation}

For aligned nuclei with $P(M_i=+I_i)=P(M_i=-I_i)=0.5$, or similarly for fully
polarized nuclei $P(M_i)=\delta_{M_i,+I_i}$, the statistical tensors
$\alpha_2 ^{I_i}$ in Eq. (\ref{stat}) are given by
$ 1/ \sqrt{5},\, \sqrt{5/14},$ and $\sqrt{7/15}$, for $I_i=3/2,\, 5/2,$ and 
$7/2$, respectively.

The transition form factors can be written in terms of the intrinsic form 
factors with the Clebsh-Gordan coefficients in Eq. (\ref{intrinsic}). The 
resulting interference form factors $F_L^{\ell=2}$ (\ref{interfff}), in terms 
of the intrinsic form factors ${\cal F}^{C0}$ and ${\cal F}^{C2}$ are given by 
\begin{equation}
  F_L^{\ell=2}(I_f;q)= {\cal A} (I_f){\cal F}^{C0}(q) {\cal F}^{C2}(q) \, ,
\end{equation}
with 
${\cal A} = -2,\, -5/7$, and $-2\sqrt{7/3}$, for 
$I_i=I_f=3/2,\, 5/2,$ and $7/2$, respectively.

\begin{figure}[bth]
\centering
\includegraphics[width=70mm]{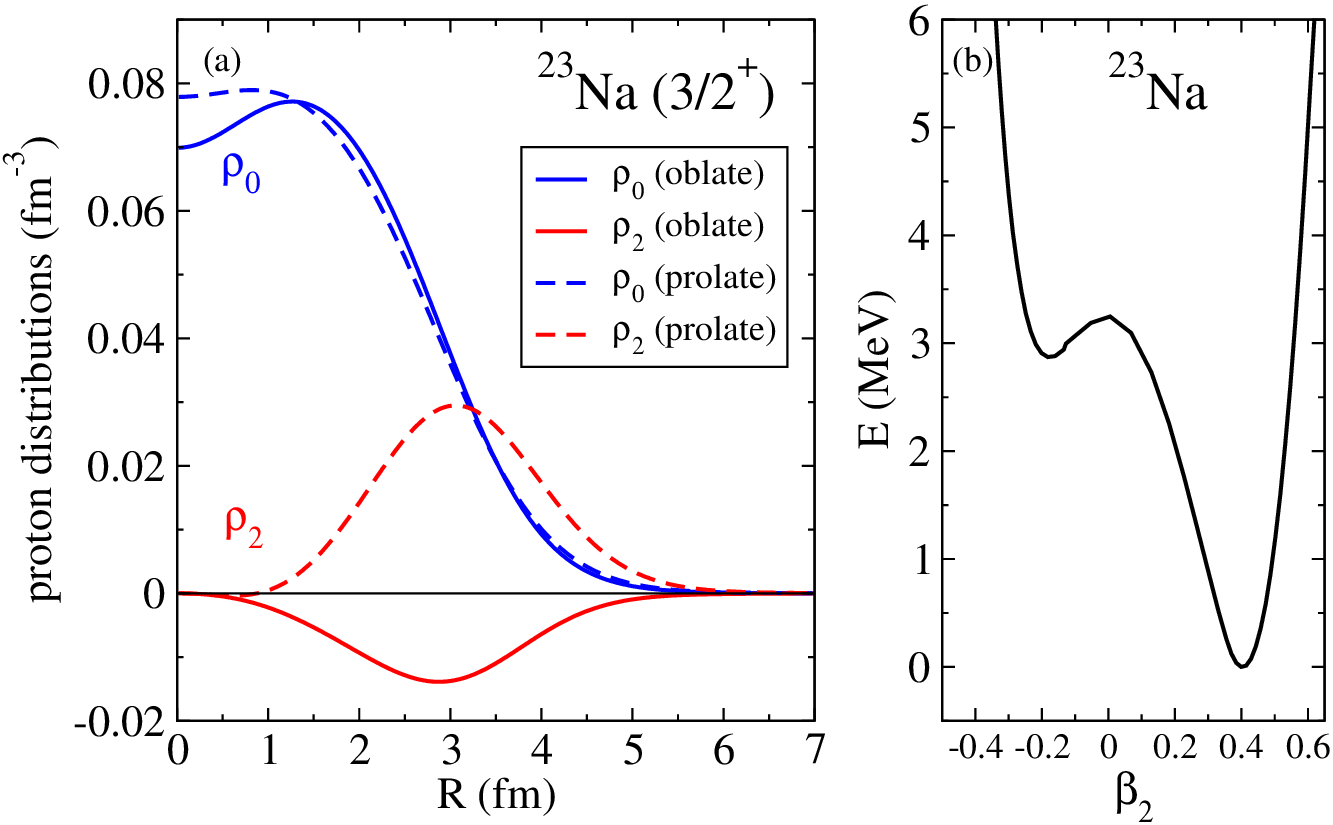}
\caption{(a) Multipole components of the proton distributions in $^{23}$Na $(I=k=3/2)$
for the oblate and prolate shape configurations. (b) Excitation energy as a function 
of the quadrupole parameter $\beta_2$.}
\label{1-23na}
\end{figure}

\begin{figure}[bth]
\centering
\includegraphics[width=70mm]{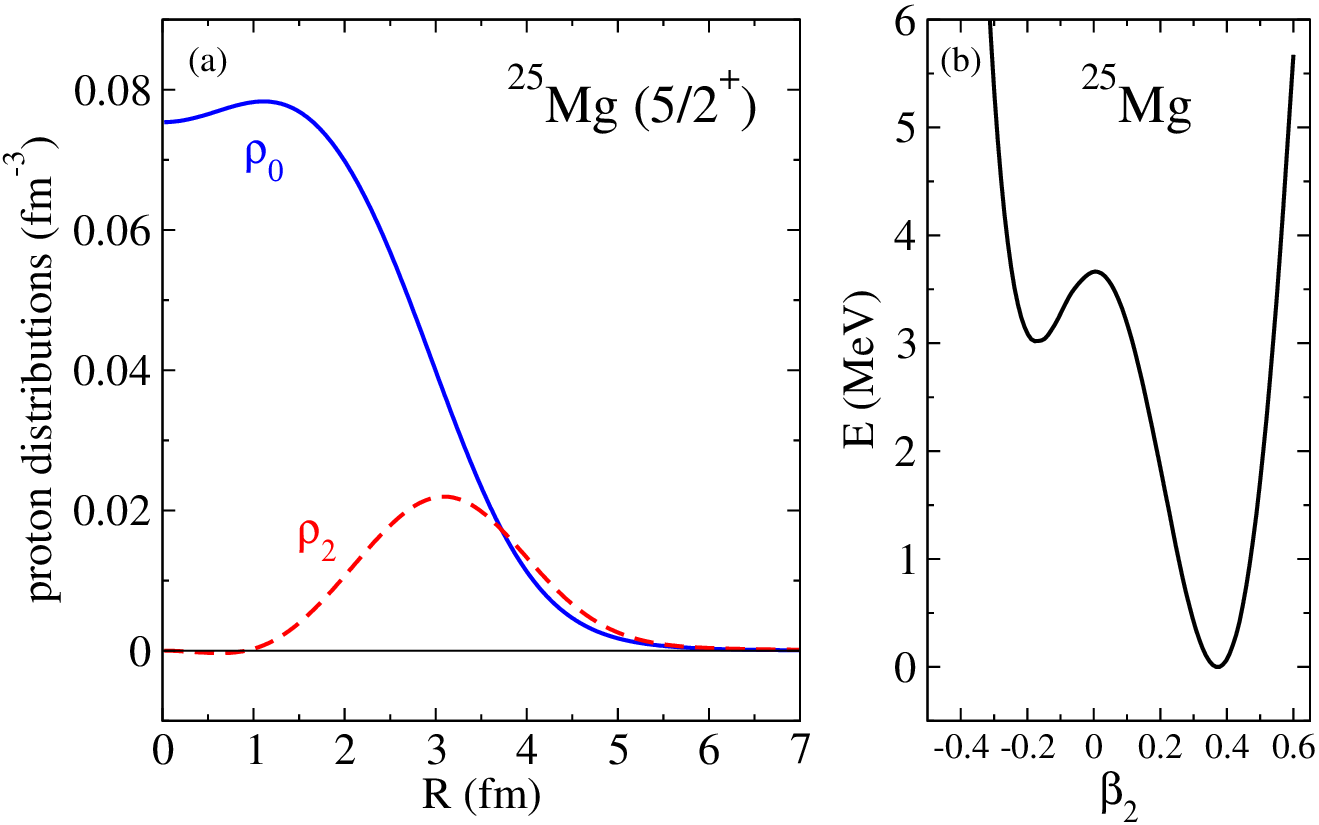}
\caption{Same as in Figure \ref{1-23na}, but for the prolate shape in $^{25}$Mg 
$(I=k=5/2)$.}
\label{1-25mg}
\end{figure}


\begin{figure}[bth]
\centering
\includegraphics[width=70mm]{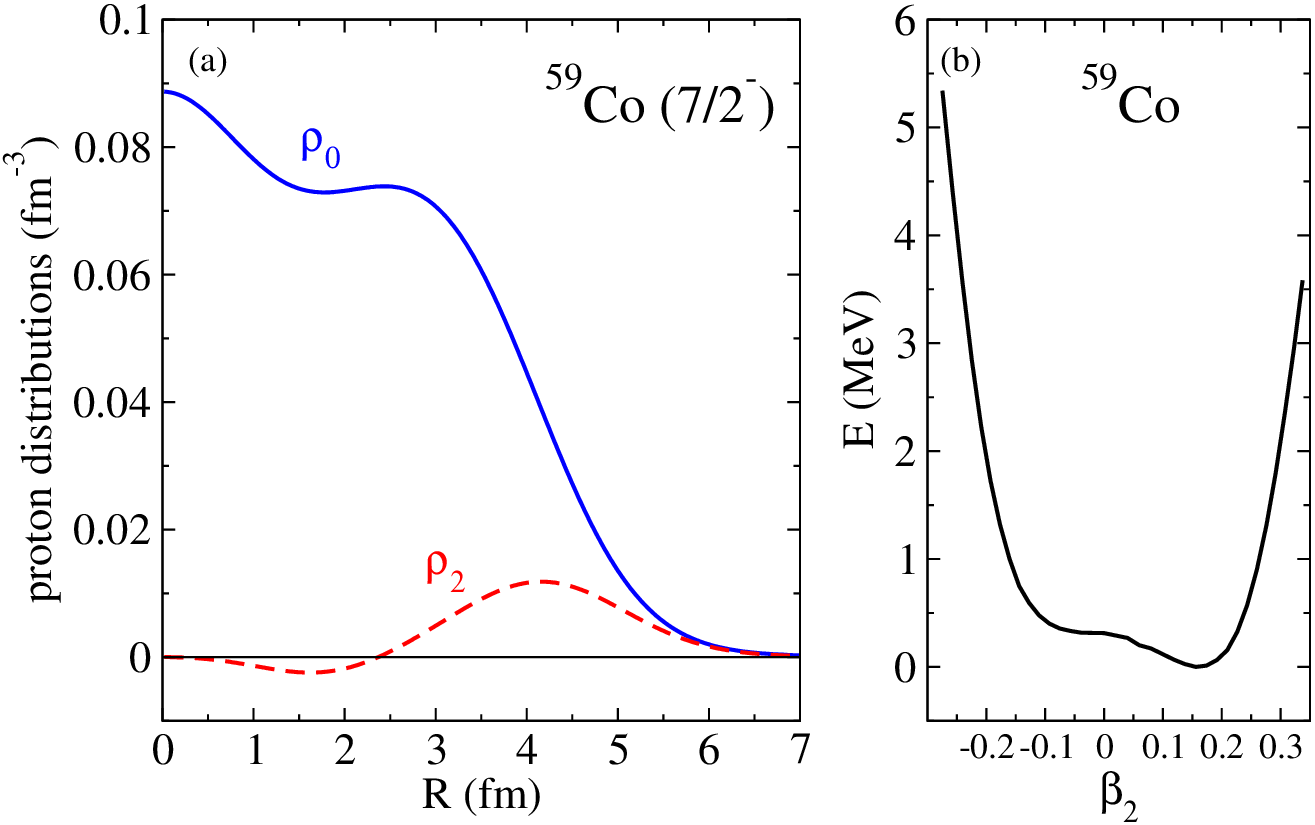}
\caption{Same as in Figure \ref{1-23na}, but for the prolate shape in $^{59}$Co 
$(I=k=7/2)$.}
\label{1-59co}
\end{figure}

\begin{figure}[bth]
\centering
\includegraphics[width=70mm]{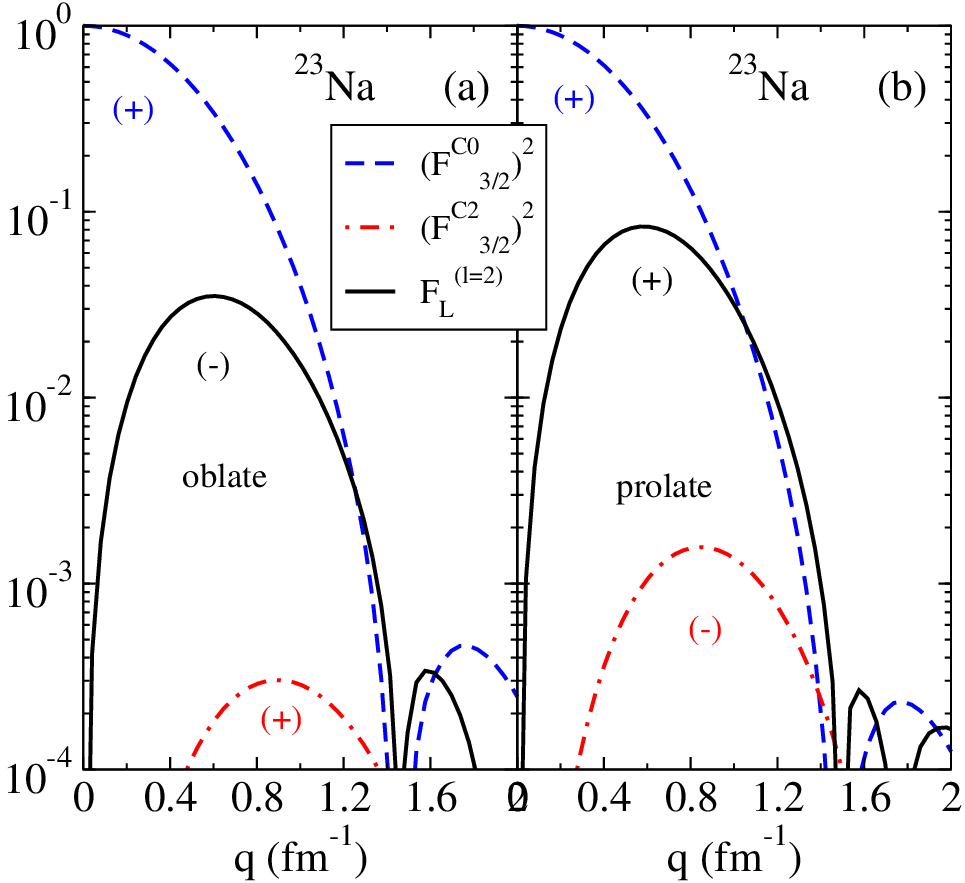}
\caption{(a) $(F^{C\lambda}_{3/2})^2$ and $F_L^{\ell =2}$ form factors for the oblate
  shape in $^{23}$Na. (b) Same for the prolate shape. Signs under the peaks in 
  $(F^{C\lambda}_{3/2})^2$ show the sign of the multipoles $F^{C\lambda}_{3/2}$.}
\label{2-23na}
\end{figure}

\begin{figure}[bth]
\centering
\includegraphics[width=70mm]{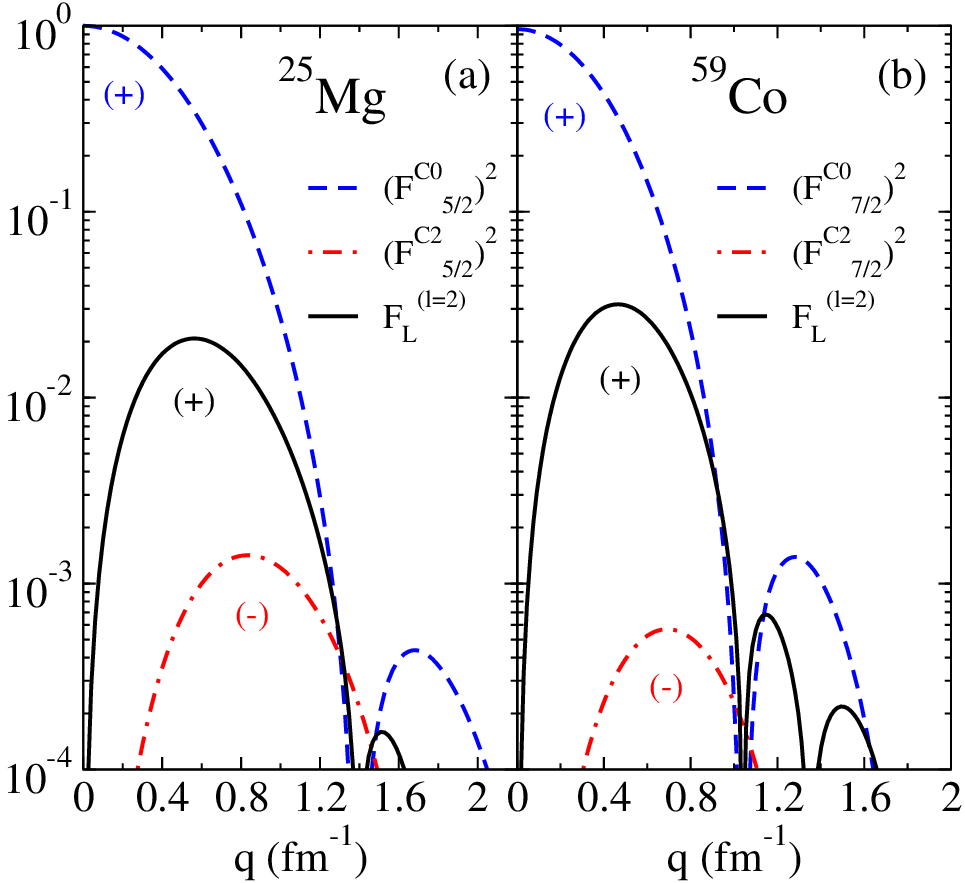}
\caption{Same as in Figure \ref{2-23na}, but for the prolate shapes of
  (a) $^{25}$Mg  ($F^{C\lambda}_{5/2}$) and (b) $^{59}$Co ($F^{C\lambda}_{7/2}$).}
\label{2-25-59}
\end{figure}

\begin{figure}[bth]
\centering
\includegraphics[width=70mm]{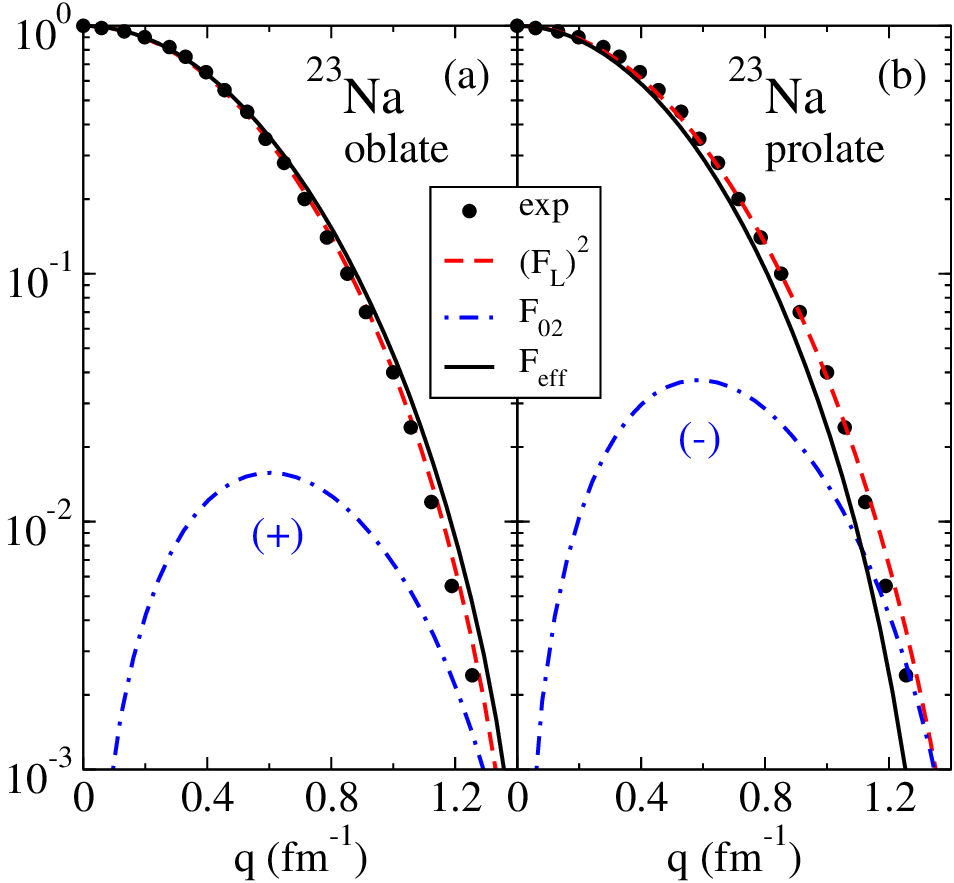}
\caption{$(F_L)^2$, $F_{02}$, and $F_{\rm eff}$ form factors in $^{23}$Na, (see text)
  for oblate (a) and prolate (b) shapes.
  Experimental data for the unpolarized form factor $(F_L)^2$ are
  taken from  \cite{liu-prc96,vries}.}
\label{3-23na}
\end{figure}

\begin{figure}[bth]
\centering
\includegraphics[width=70mm]{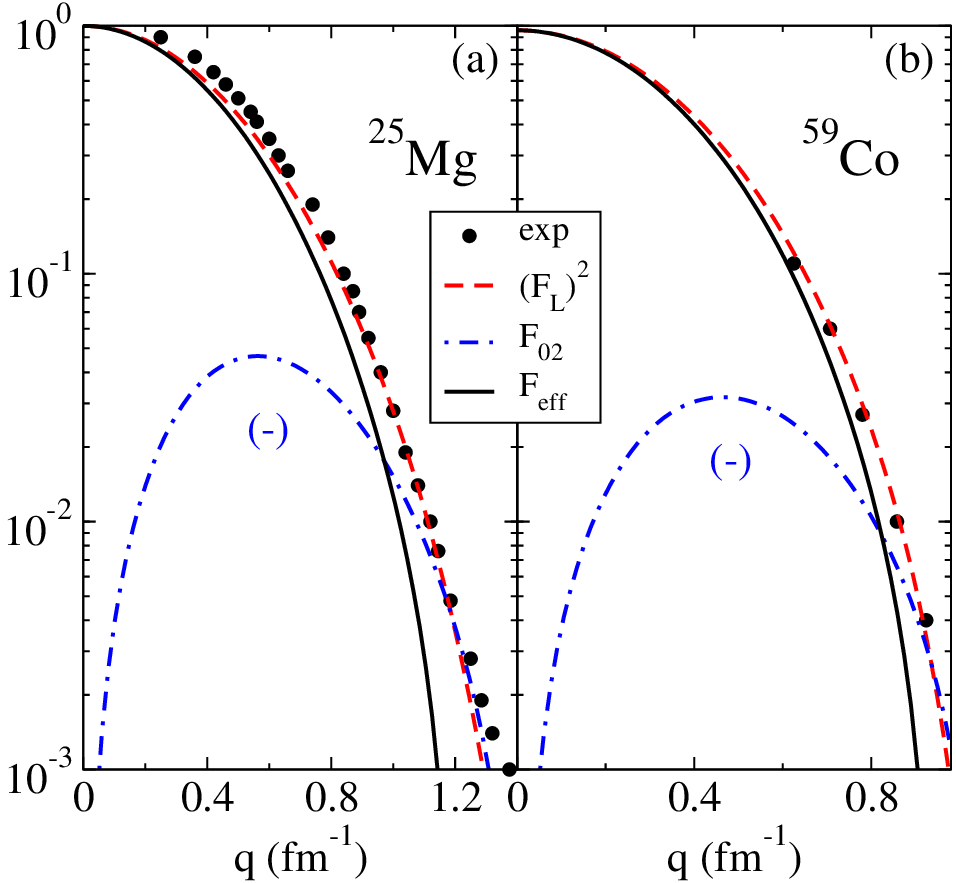}
\caption{Same as in Figure \ref{3-23na}, but for the prolate shapes in $^{25}$Mg (a)
  and $^{59}$Co (b).
Experimental data are taken from \cite{liu-prc96,exp25mg,exp59co}.}
\label{3-25-59}
\end{figure}

\begin{figure}[bth]
\centering
\includegraphics[width=70mm]{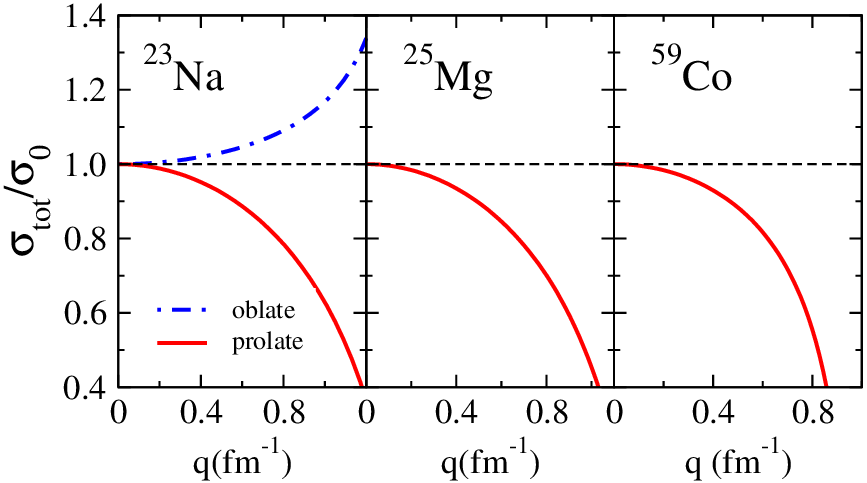}
\caption{Ratio between the total cross-section with aligned nuclei and 
the unpolarized one.}
\label{4}
\end{figure}

\section{Results}

In this section, the results for the multipole form factors and interferences 
are presented on the example of some selected deformed and stable nuclei. 
Specifically, we study $^{23}$Na ($I^{\pi}_{\rm g.s.}=3/2^+$), $^{25}$Mg 
($I^{\pi}_{\rm g.s.}=5/2^+$), and  $^{59}$Co ($I^{\pi}_{\rm g.s.}=7/2^-$).

Figures \ref{1-23na}, \ref{1-25mg}, and \ref{1-59co} contain the results for 
the multipole densities $\rho_0$ and $\rho_2$ for the proton distributions in 
$^{23}$Na, $^{25}$Mg, and $^{59}$Co, respectively, as well as the ground-state
energies as a function of the quadrupole deformation parameter $\beta_2$,
defined in Eq. (\ref{beta2}).
In the case of $^{23}$Na, both prolate and oblate deformations are considered for
the calculations since both minima correspond to $3/2^+$ states in agreement with the
experimental assignment. In the other two cases only the
prolate shape is studied because the spin-parities match the experiment,
$5/2^+$ in $^{25}$Mg and $7/2^-$ in $^{59}$Co, while the oblate minima
correspond to different states.
From these figures one can see that $\rho_2(R)$ mainly peaks in the surface 
region, being positive or negative depending on the sign of deformation. On 
the other hand, the density does not change much in the nuclear interior.
This sign difference will manifest itself in the $C2$ form factors becoming
a signature of the  oblate/prolate character of the nuclear shape.

The next figures, \ref{2-23na} and \ref{2-25-59}, contain
the Coulomb form factors squared $(F^{C0}_{I_f})^2$, $(F^{C2}_{I_f})^2$, and
$F_L^{\ell=2}$. Figure \ref{2-23na} shows the results for the oblate and
prolate shapes in $^{23}$Na, while figure \ref{2-25-59} shows the results
for the prolate shape in $^{25}$Mg and $^{59}$Co. Plus or minus signs
under the peaks of the form factors squared indicate the sign of the
corresponding multipole form factor  $F^{C\lambda}_{I_f}$.

From these figures one can 
appreciate the obvious difficulty to extract information on $C2$ multipoles 
from unpolarized experiments. First, because of their small magnitude 
compared to $C0$, which means that they only manifest significantly at 
diffraction minima. Secondly, because only the squared $(F^{C2})^2$ are 
measurable and therefore, no information about the sign of $\rho_2$ can 
be extracted. On the other hand, the interference term $F_L^{\ell=2}$ is much 
larger than $(F^{C2})^2$ and contains information of the sign of deformation.
Since the terms $F_L^{\ell=2}$ can be experimentally isolated, their comparison
with the usual $(F^{C0})^2$ and $(F^{C2})^2$  form factors is of great interest.
Typically, the first peak of $F_L^{\ell=2}$ is about one order of magnitude 
smaller than  $(F^{C0})^2$ and one order of magnitude larger than  $(F^{C2})^2$.

Under the conditions established in Section 2, the longitudinal contribution 
to the cross-section can be written as 

\begin{equation}
 \sigma_{\rm tot}= \sigma_0 + \sigma_{\rm al} \sim V_L F_{\rm eff}\, ,
\end{equation}
where

\begin{equation}
  F_{\rm eff} =  |F_L|^2 + F_{02} =
  |F^{C0}|^2 + |F^{C2}|^2 + F_{02} \, ,
\end{equation}
with

\begin{equation}
F_{02} = - \alpha_2^{I_i} P_2(\cos \theta') F_L^{(\ell=2)} \, .
\end{equation}

Figures \ref{3-23na} and \ref{3-25-59} contain
the total calculated and measured longitudinal form factor 
$(F_L)^2=(F^{C0})^2+(F^{C2})^2$ in the unpolarized case,
together with the above defined $F_{02}$ and $F_{\rm eff}$ form factors 
for $^{23}$Na oblate and prolate and for prolate shapes in $^{25}$Mg and $^{59}$Co.
Comparing the unpolarized response, which is proportional to $(F_L)^2$
with the polarized one, which is proportional to $F_{\rm eff}$, one
can see that simply measuring both the unpolarized and polarized
cross-sections and checking whether it increases or decreases,
information on the oblate or prolate type of the nuclear shape
can be inferred.
This relative comparison of cross-sections would give us a model 
independent signature of the sign of deformation. Further isolation
of  $F_L^{(\ell=2)}$ will provide more detailed information about $\rho_2$.

Figure \ref{4} shows the ratio between the total polarized cross-section 
with aligned nuclei and the unpolarized one,

\begin{equation}
\frac{\left( \sigma_0 + \sigma_{\rm al} \right)}{\sigma_0} = 
1+ \frac{F_{02}}{|F_L|^2} = \frac{F_{\rm eff}}{|F_L|^2} \, .
\end{equation}
It clearly shows the magnitude of the deviation from the standard unpolarized
cross-section. The change begins to be significant already at low momentum 
transfer and becomes sizable below the first diffraction minimum, where the 
cross-section is large enough to be measured. In the neighborhood of the 
minimum the cross-section is very small and other effects not taken into 
account in this study may appear.

\section{Conclusions}

The effect of deformation on the interference form factors between
monopole and quadrupole Coulomb terms, and therefore, on the total 
electron-scattering cross-section has been studied in odd-$A$ nuclei.
Examples of the magnitude of the expected effects are shown in specific 
cases with different spin-parity ground states, namely, $^{23}$Na ($I^{\pi}=3/2^+$), 
$^{25}$Mg ($I^{\pi}=5/2^+$), and  $^{59}$Co ($I^{\pi}=7/2^-$).
In this work, the focus is placed on elastic scattering with unpolarized 
electrons, but with oriented odd-$A$ axially deformed nuclei. The reaction 
mechanism is based on the PWBA, whereas the nuclear structure is described 
with selfconsistent Skyrme HF+BCS calculations.

A more systematic and general study of the interferences between different
multipoles, without the restrictions made in this work will be performed
in a forthcoming publication \cite{next}, where issues like inelastic 
reactions, polarized electrons, higher spins, Coulomb distortions, or
interferences other than $C0/C2$ will be considered.
It will be also interesting to study the sensitivity of the cross-section
to the nuclear shape in nuclei exhibiting shape coexistence, where oblate
and prolate configurations have similar energies.
The present study provides clues
about the different effects expected from oblate or prolate shapes.

The limitations imposed in this work do not alter the final 
conclusions regarding the possibility of measuring observables, which are 
sensitive to the $C0/C2$ interference and, as a consequence, are sensitive to
the sign of the axial deformation in odd-A nuclei.
This model independent signature appears as an increase or decrease of 
the unpolarized cross-section. The magnitude of such
effect will depend on the particular nuclear model used.


\section*{Acknowledgments}

This work was supported by Grant PGC2018-093636-B-I00, funded by 
MCIN/AEI/10.13039/501100011033 and by ERDF 'A way of making Europe'.





\end{document}